\documentclass[aps,prl,amsmath,floats,floatfix,twocolumn,
  superscriptaddress,nofootinbib,showpacs]{revtex4-1}

\pdfoutput=1 
\usepackage{diagbox}
\usepackage{mathrsfs}
\usepackage{multirow}
\usepackage{braket}
\usepackage{amsfonts}
\usepackage{amsmath}
\usepackage{amssymb}
\usepackage{amsthm}
\usepackage{bm}
\usepackage{enumitem}
\usepackage{graphicx}
\usepackage{graphics}
\usepackage{latexsym}
\usepackage{rotating}
\usepackage[dvipsnames]{xcolor}
\usepackage{mathrsfs}
\usepackage{microtype}
\usepackage{verbatim}
\usepackage{url}

\usepackage[caption=false]{subfig}
\usepackage[normalem]{ulem}

\usepackage[breaklinks=true]{hyperref}
\hypersetup{
    colorlinks=true,
    linkcolor=NavyBlue,
    filecolor=Magenta,
    urlcolor=NavyBlue,
    citecolor=NavyBlue
}

\usepackage{yfonts}
\usepackage{float}
\usepackage{xspace}
\usepackage{mathrsfs}
\usepackage{siunitx}
\usepackage{array}

\allowdisplaybreaks[4]

\newcommand{\be}{\begin{equation}}
\newcommand{\ee}{\end{equation}}

\newcommand{\ba}{\begin{align}}
\newcommand{\ea}{\end{align}}

\makeatletter
\newcommand*{\rom}[1]{\expandafter\@slowromancap\romannumeral #1@}
\makeatother

\AtBeginDocument{%
    \newwrite\bibnotes
    \def\bibnotesext{Notes.bib}
    \immediate\openout\bibnotes=\jobname\bibnotesext
    \immediate\write\bibnotes{@CONTROL{REVTEX41Control}}
    \immediate\write\bibnotes{@CONTROL{%
    apsrev41Control,author="08",editor="1",pages="1",title="0",year="1"}}
    \if@filesw
    \immediate\write\@auxout{\string\citation{apsrev41Control}}%
    \fi
}

\allowdisplaybreaks

\begin{document}

\title{Measuring a Black Hole's Area Immediately after Merger: \\ A Direct-Wave Test of Hawking's Area Law}

\author{Adrian Ka-Wai Chung}
\email{kwc43@cam.ac.uk}
\affiliation{DAMTP, Centre for Mathematical Sciences, University of Cambridge,
Wilberforce Road, Cambridge CB3 0WA, United Kingdom}

\author{Kelvin Ka-Ho Lam}
\affiliation{Illinois Center for Advanced Studies of the Universe \& Department of Physics,
University of Illinois Urbana-Champaign, Urbana, Illinois 61801, USA}

\author{Anna Liu}
\affiliation{Illinois Center for Advanced Studies of the Universe \& Department of Physics,
University of Illinois Urbana-Champaign, Urbana, Illinois 61801, USA}

\author{Nicol\'as Yunes}
\affiliation{Illinois Center for Advanced Studies of the Universe \& Department of Physics,
University of Illinois Urbana-Champaign, Urbana, Illinois 61801, USA}


\begin{abstract}
Black-hole area is the geometric variable behind horizon thermodynamics. We introduce a gravitational-wave method to infer a Kerr-equivalent horizon area from direct waves in the near-merger signal, before quasinormal ringing dominates at late times. Applied to GW250114, and interpreting the fitted direct-wave frequency and damping rate as horizon quantities, we find that analyses initiated $3$--$4.5M$ before the peak-amplitude time yield an area consistent with the Kerr remnant. This result gives a first area measurement using direct waves and a new near-merger test of Hawking's area law.

\end{abstract}

\maketitle

\noindent \textit{Introduction.} The event-horizon area (or ``horizon area,'' for short) of a black hole is a fundamental quantity that characterizes its geometry and physical properties.
The horizon area of a black hole is directly related to its entropy \cite{Bekenstein:1973ur}, which underlies the concepts of black-hole temperature and Hawking radiation \cite{Hawking:1974rv, Hawking:1975vcx}.
Beyond its thermodynamic interpretation, the horizon area provides a compact geometric characterization of a black hole's mass, spin, and energy balance~\cite{Mino_2008, Zimmerman_2011}, and may offer a useful observable for population inference~\cite{LIGOScientific:2025pvj} and for studies of near-horizon and accretion dynamics.
Given its importance, developing observational methods to measure the horizon area is of significant interest.

Electromagnetic observations can probe horizon-scale geometry~\cite{Blandford:1977ds, event2019first}, but, in practice, such inferences are limited by emission physics and surrounding accretion flows.
In contrast, gravitational-wave observations of black-hole coalescences in the $\sim 10$–$200 M_{\odot}$ mass range offer a cleaner probe, as they are typically not accompanied by luminous accretion.
Binary black-hole mergers involve highly dynamical horizon evolution and constitute the dominant source of detected gravitational waves \cite{LIGOScientific:2025slb, LIGOScientific:2021djp, Abbott:2020niy, LIGO_10}, making them particularly powerful systems for horizon area measurements. 
Previous gravitational-wave measurements of horizon area~\cite{Isi:2020tac, LIGOScientific:2025rid} inferred the progenitor areas from the inspiral and the remnant area from the late-time, approximately linear quasinormal-mode ringdown. These measurements were designed primarily as tests of Hawking's area law, which states that the total black-hole horizon area cannot decrease in classical general relativity~\cite{Hawking:1971tu, Bardeen:1973gs}. 
Applied to GW250114, this strategy showed that the remnant horizon area is larger than the sum of the progenitor areas~\cite{LIGOScientific:2025rid, LIGOScientific:2025wao}.

\begin{figure}[tp!]
\centering  
\subfloat{\includegraphics[width=\columnwidth]{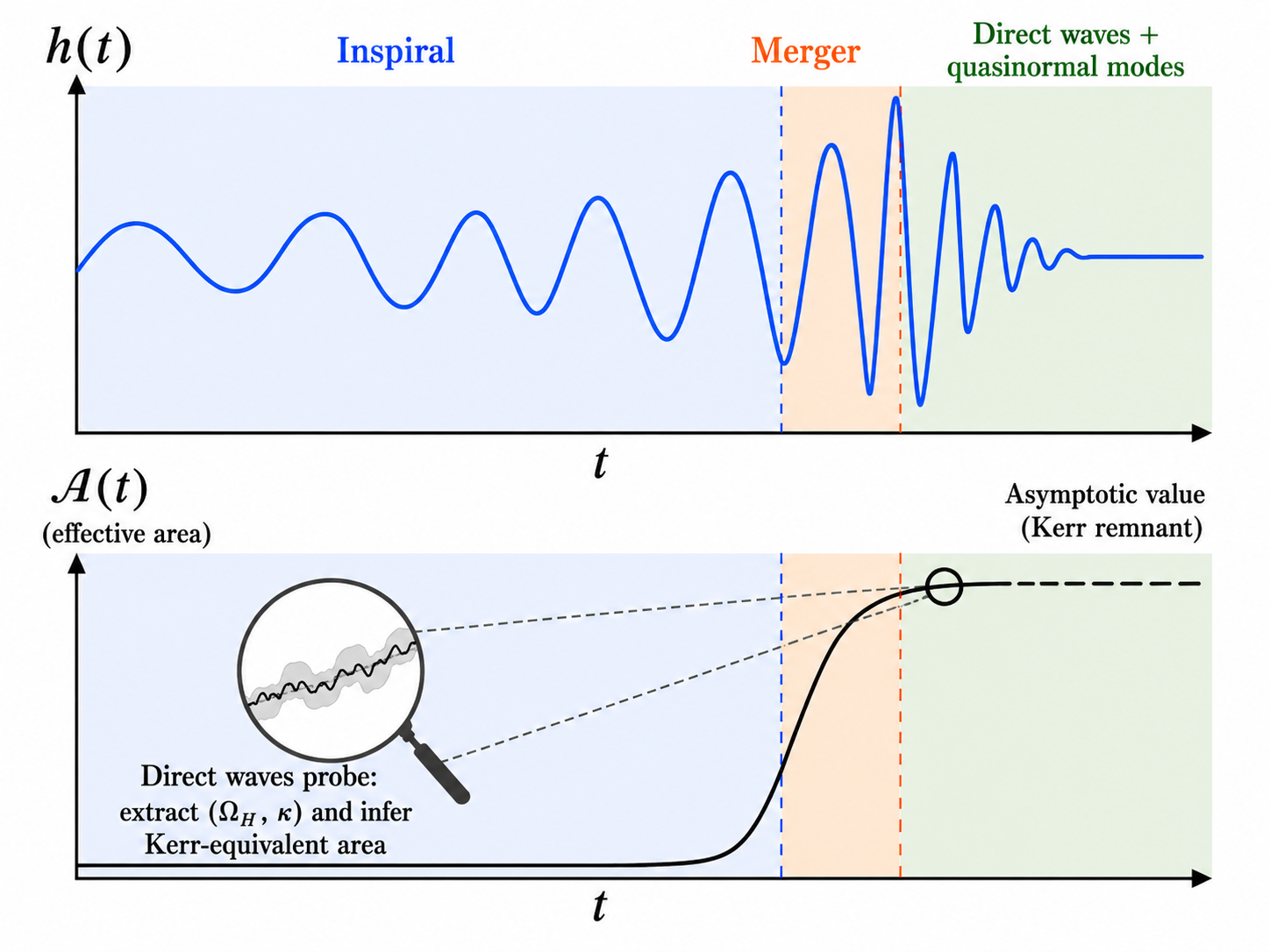}}
\caption{Schematic illustration of the direct-wave measurement of black-hole horizon area. 
The top panel shows the gravitational-wave strain $h(t)$ across the inspiral, merger, and post-merger stages. 
The orange band indicates the near-peak interval used to search for direct waves, while the green band denotes the regime containing direct-wave and quasinormal-mode contributions. 
The bottom panel shows the corresponding effective Kerr-equivalent area $\mathscr{A}(t)$. 
Direct waves provide estimates of $(\Omega_{\rm H},\kappa)$, which are mapped to a Kerr-equivalent horizon area, while the late-time ringdown determines the asymptotic Kerr-remnant area. 
The magnified region emphasizes the time-localized nature of the direct-wave probe.
}
\label{fig:schematic}
\end{figure}

In this work, we present a new method to measure the black-hole horizon area from gravitational-wave data by exploiting direct-wave signals in the near-merger portion of binary black-hole coalescence (see Fig.~\ref{fig:schematic}).
These signals, recently identified in numerical and analytical studies~\cite{Kankani:2026byb, Lu:2025vol, Oshita:2025qmn}, arise before the waveform has settled into the late-time, linear quasinormal-mode regime.
In the GW250114 regime studied here, their fitted frequency and damping rate can be interpreted as estimates of the remnant horizon angular velocity and surface gravity.
This mapping yields an estimate of the remnant horizon area during the merger-ringdown transition, providing a time-localized probe that complements existing methods based on the inspiral and late-time ringdown.

\vspace{0.2cm}

\noindent \textit{Review of the Horizon Area and Area Law.} To frame our measurement of the horizon area in a logically consistent framework, we explicitly state the underlying assumptions:
\begin{enumerate}[label=A.\arabic*,start=1]
    \item \label{item:A1} General relativity is correct, matter satisfies the null energy condition, spacetime is strongly asymptotically predictable and contains an event horizon, and quantum effects, such as Hawking radiation, can be ignored.
    \item \label{item:A2} A stationary Kerr black hole has horizon area
    \begin{equation}\label{eq:A_RD}
        \mathscr{A} = 8 \pi M^2 \left(1+\sqrt{1-a^2}\right),
    \end{equation}
    where $M$ and $a$ are the mass and dimensionless spin of the black hole, respectively.
    \item \label{item:A3} For perturbations between neighboring stationary black-hole spacetimes, the first law of black-hole mechanics,
    \begin{equation}
    \label{eq:first-law}
        \delta \mathscr{A}
        =
        \frac{8\pi}{\kappa}
        \left(\delta E-\Omega_{\rm H}\delta J\right),
    \end{equation}
    is valid, where $\delta E$ and $\delta J$ are first-order changes in the black hole's mass-energy and angular momentum, and $\kappa$ and $\Omega_{\rm H}$ are the surface gravity and horizon angular velocity of the unperturbed black hole.
\end{enumerate}

These assumptions are standard, but not independent of each other. Assumption~\ref{item:A1} specifies the classical setting in which Hawking's area theorem applies~\cite{Hawking:1971tu,Hawking:1973uf,Bardeen:1973gs}. Assumption~\ref{item:A2} follows from the Kerr solution and black-hole uniqueness theorems of general relativity~\cite{Carter:1968rr,Israel:1967za,Robinson:1975bv}. Assumption~\ref{item:A3} is the first law of black-hole mechanics, originally derived by Bardeen, Carter, and Hawking~\cite{Bardeen:1973gs}; see also Refs.~\cite{Wald:1984cw,Poisson:2009pwt}. Under assumption~\ref{item:A1}, Hawking's area law states that the total event-horizon area is non-decreasing in classical general relativity, $\delta \mathscr{A} \geq 0$.

How are these concepts related to the horizon area used in our work? We are concerned with the merger and immediate post-merger phases of black-hole binaries, so we do not assume that the exact spacetime during the merger-ringdown transition is stationary or axisymmetric. Instead, we use Eq.~\eqref{eq:A_RD} to define a Kerr-equivalent ``effective area,'' associated with the remnant horizon quantities inferred from the direct-wave and ringdown regimes. This is not a reconstruction of the full nonlinear event-horizon area. Rather, it is a leading-order remnant quantity whose interpretation becomes valid only once the post-merger spacetime is sufficiently well described as a perturbed Kerr black hole. We therefore \textit{test} this interpretation \textit{a posteriori} by varying the direct-wave analysis start time and identifying the regime in which the inferred effective area is consistent with the independently measured ringdown area and with the Kerr remnant prediction. Under this interpretation, it is consistent to compare the Kerr-equivalent areas inferred from the direct-wave and ringdown regimes. 

A rigorous treatment of the exact dynamical horizon area would require horizon tracking in numerical relativity or a dynamical-horizon framework~\cite{Ashtekar:2003hk,Ashtekar:2004cn}. In such a framework, the area is the quasi-local area of marginally trapped surfaces, and its change is governed by exact flux-balance laws. These laws require local horizon data, such as the horizon shear, connection terms, angular-momentum density, and, in rotating cases, a preferred axial vector field. Such quantities can be computed in numerical-relativity simulations with horizon tracking, but they are not determined by the two direct-wave fit parameters $\Omega_{\rm H}$ and $\kappa$ alone.

\vspace{0.2cm}

\noindent \textit{Methods.}~Theoretical calculations and numerical simulations suggest that gravitational waves from the late inspiral and merger phase onward can be modeled as a sum of a direct wave and a ringdown signal, the latter consisting of a discrete set of quasinormal modes. 
This picture allows us to build the following phenomenological time-domain waveform model, 
\begin{align}
\label{eq:full-model}
h(t) &= h_{+}(t) - i h_{\times} (t) = \Theta(t-t_{0}) \left[ h_{\rm DW}(t) + h_{\rm RD}(t) \right], 
\end{align}
where $\Theta$ restricts the model to the analysis window $t\geq t_0$, with $t_0$ the start time that we vary below as a robustness check. Here, $h_{\rm DW}(t)$ denotes the direct-wave component, and $h_{\rm RD}(t)$ denotes the ordinary, linear quasinormal-mode ringdown component. The latter depends on the mass and spin of the late-time stationary Kerr remnant, and its explicit expression is given, for example, in Eq.~(1) of Ref.~\cite{LIGOScientific:2026wpt}.

The direct-wave component can be approximated as \cite{Kankani:2026byb, Lu:2025vol, Oshita:2025qmn}
\begin{equation}
    h_{\rm DW}(t) \!\approx\! B_{\rm DW} \!
\exp\left[-\gamma_{\rm DW}(t-t_0)\!+\!i\omega_{\rm DW}(t-t_0)\!+\!i\phi_{\rm DW}\right],
\end{equation}
where, $B_{\rm DW}$, $\gamma_{\rm DW}$, $\omega_{\rm DW}$, and $\phi_{\rm DW}$ denote the amplitude, damping rate, angular frequency, and phase of the direct waves, respectively. 

In the effective horizon interpretation used below, \(\gamma_{\rm DW}\simeq\kappa\) and \(\omega_{\rm DW}\simeq m\Omega_{\rm H}\), so that \(\Omega_{\rm H}=\omega_{\rm DW}/m\) when constructing \(\mathscr A_{\rm DW}\), where $m$ is the azimuthal mode number.
These parameters are fitted phenomenological quantities associated with the orbital motion of the effective near-horizon perturber, and are therefore not assumed to be identically equal to $m\Omega_{\rm H}$ and $\kappa$ at all times.
The effective horizon interpretation used in this work assumes that, sufficiently close to merger and during the merger-ringdown transition, the perturber has been strongly frame-dragged and redshifted so that($\omega_{\rm DW}\simeq m\Omega_{\rm H}$ and $\gamma_{\rm DW}\simeq\kappa$.
We test the validity of this interpretation \textit{a posteriori}, by varying the analysis start time ($t_0$) and identifying the regime in which the inferred Kerr-equivalent area agrees with the independently inferred ringdown area.
Analytical calculations and numerical simulations suggest that, when the post-merger spacetime is sufficiently well described as a perturbed Kerr black hole, the fitted direct-wave frequency and damping rate can be interpreted as effective estimates of the horizon angular velocity and surface gravity of the Kerr background.

For a stationary Kerr black hole of mass $M$ and dimensionless spin $a$, these quantities are
\begin{align}
\label{eq:OmegaH-and-Kappa} 
\Omega_{\rm H} &= \frac{a}{2M\left(1+\sqrt{1-a^2}\right)}, \quad
\kappa = \frac{\sqrt{1-a^2}}{2M\left(1+\sqrt{1-a^2}\right)}\,.
\end{align}
In the direct-wave regime, we therefore treat the post-merger black hole, in an effective sense, as a perturbed Kerr black hole whose fitted $\Omega_{\rm H}$ and $\kappa$ define Kerr-equivalent horizon quantities.
This identification is not assumed to hold at all start times; it is tested below by varying $t_0$.
Under assumption~\ref{item:A2}, the fitted $\Omega_{\rm H}$ and $\kappa$ can then be used to infer an effective Kerr horizon area by eliminating $M$ and $a$ from Eqs.~\eqref{eq:A_RD}--\eqref{eq:OmegaH-and-Kappa}.
Expressed directly in terms of $\Omega_{\rm H}$ and $\kappa$, this yields
\begin{equation}\label{eq:A_DW}
\mathscr{A}_{\rm DW} = \frac{2 \pi}{\sqrt{\Omega_{\rm H}^2 + \kappa^2}\left(\sqrt{\Omega_{\rm H}^2 + \kappa^2}+\kappa \right)}.
\end{equation}

Using this waveform model, we estimate $\mathscr{A}_{\rm DW}$ and $\mathscr{A}_{\rm RD}$ with two complementary analyses:
\begin{enumerate}[label=M.\arabic*,start=1]
    \item \label{item:S1} For a given analysis start time $t_0$, measured relative to the peak-amplitude time of the dominant $(2,2)$ mode, we infer the posterior distributions of $B_{\rm DW}$, $\gamma_{\rm DW}$, $\omega_{\rm DW}$, and $\phi_{\rm DW}$, together with the amplitudes and phases of the $(n\ell m)=(022,122)$ quasinormal modes, and the luminosity distance. In this step, we fix the detector-frame mass and dimensionless spin of the late-time stationary Kerr remnant to $M = M_{\rm f}^{\rm det}$ and $a$, using the maximum-posterior values obtained from an inspiral--merger--ringdown analysis.
    \item \label{item:S2} At a late start time, $t_0 = +10M_{\rm f}^{\rm det}$, measured relative to the same peak-amplitude time, we perform a standard ringdown spectroscopy analysis with $h_{\rm DW}(t)\equiv 0$ to infer the mass and spin of the late-time Kerr remnant.
\end{enumerate}
Method~\ref{item:S1} yields estimates of $\mathscr{A}_{\rm DW}$ via Eq.~\eqref{eq:A_DW} and the posterior estimates of $\Omega_{\rm H}$ and $\kappa$.
In this method, fixing the late-time Kerr-remnant parameters $(M,a)$ is necessary to avoid strong degeneracies among $(M,a)$, $\Omega_{\rm H}$, and $\kappa$ in the waveform model.
This choice follows the strategy adopted in Ref.~\cite{Lu:2025vol}.
It does not by itself impose equality between $\mathscr{A}_{\rm DW}$ and $\mathscr{A}_{\rm RD}$: the direct-wave area is computed from the fitted direct-wave parameters, while $\mathscr{A}_{\rm RD}$ is inferred independently in Method~\ref{item:S2}, together with the late-time remnant mass and dimensionless spin. A full marginalization over the late-time remnant parameters is deferred to future work.

In Method~\ref{item:S1}, both the direct-wave and ringdown components of the model are activated from the same start time $t_0$.
This should be understood as a fitting prescription: the ringdown term models the quasinormal-mode contribution present in the same analysis window, while the direct-wave term captures the remaining near-merger contribution.
In this sense, Method~\ref{item:S1} effectively fits and subtracts the modeled quasinormal ringdown signal, based on fixed late-time remnant parameters, from the data, following Ref.~\cite{Lu:2025vol}.

\vspace{0.2cm}

\noindent \textit{Parameter Estimation.} We apply the above formalism to measure $\mathscr{A}_{\rm DW}$ and $\mathscr{A}_{\rm RD}$ from GW250114, the first signal reported to exhibit a direct-wave component \cite{Lu:2025vol}.
By using the quasi-circular, spin-precessing \texttt{NRSur7dq4} model, the best-fit, detector-frame, final (late-time remnant) mass was inferred to be $M_{\rm f}^{\rm det} = 68.1\,M_{\odot}$ and the corresponding final spin $a = 0.68$ \cite{LIGOScientific:2025rid, LIGOScientific:2025wao}. 
We here re-analyze the strain data measured by the Hanford and Livingston detectors~\cite{LIGOScientific:2014pky} with a duration of 0.2 seconds, starting from $t_0$ before the merger time of GW250114 (i.e.~before the peak-amplitude of the 22-mode of the signal).
We perform nested sampling using \texttt{pyRing} \cite{pyRing}, a python package for time-domain black-hole ringdown spectroscopy. 
In both methods 1 and 2, the $(nlm) = (022) $ and (122) modes are included, as the overtone $(nlm)=(122)$ mode was reported to have been detected~\cite{abac2026black}.
The posterior is sampled with 4096 live points.  Each posterior sample of $(\Omega_{\rm H}, \kappa)$ and $(M, a)$ is converted into a sample of $\mathscr{A}_{\rm DW}$ and $\mathscr{A}_{\rm RD}$. 

\begin{figure}[tp!]
\centering  
\subfloat{\includegraphics[width=\columnwidth]{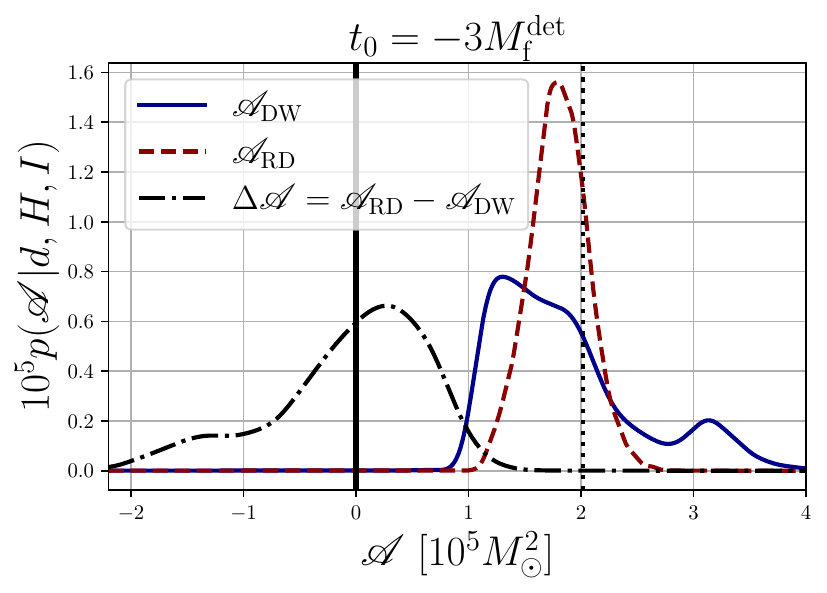}}
\caption{The probability distribution function of the horizon area during the emission of direct waves ($\mathscr{A}_{\rm DW}$, solid blue), ringdown phase ($\mathscr{A}_{\rm RD}$, dashed red), and the area increment ($\Delta \mathscr{A} = \mathscr{A}_{\rm RD} - \mathscr{A}_{\rm DW}$) obtained by searching for the direct waves of the GW250114 event.
The analysis starts at $t_0 = -3M_{\rm f}^{\rm det}$, where $M_{\rm f}^{\rm det}$ is the detector-frame remnant mass corresponding to the maximum-posterior estimate from an initial full inspiral-merger-ringdown signal inference (see main text for details). 
The negative start time indicates that the search begins prior to the merger. 
The dashed vertical line marks the horizon area computed from the maximum-posterior remnant mass and dimensionless spin. 
}
\label{fig:PDF_A_m3M}
\end{figure}

Figure~\ref{fig:PDF_A_m3M} shows the posterior distributions of $\mathscr{A}_{\rm DW}$ (solid blue) and $\mathscr{A}_{\rm RD}$ (dashed red) in units of $M_{\odot}^2$.
The posterior of $\mathscr{A}_{\rm DW}$ is obtained from a direct-wave search with starting time $t_0 = -3 M_{\rm f}^{\rm det}$, corresponding to the latest start time considered in \cite{Lu:2025vol}.
For comparison, the vertical dashed line marks the predicted horizon area computed from $M_{\rm f}^{\rm det} = 68.1 M_{\odot}$ and $a = 0.68$. 
Under assumption~\ref{item:A2}, this predicted value should be consistent with both $\mathscr{A}_{\rm DW}$ and $\mathscr{A}_{\rm RD}$.
We find that both posteriors exhibit significant support near the predicted horizon area, indicating that both the direct-wave analysis and ringdown spectroscopy are capable of estimating the horizon area.

\begin{figure}[tp!]
\centering  
\subfloat{\includegraphics[width=\columnwidth]{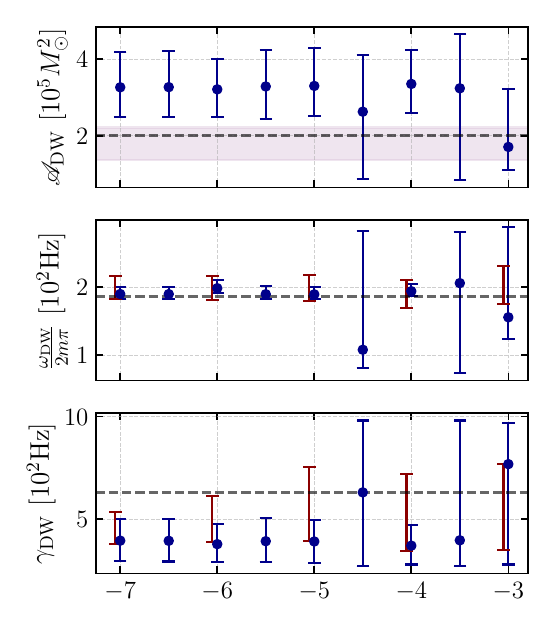}}
\caption{The median (markers) and 90\% credible intervals (error bars) of the marginalized posteriors for $\mathscr{A}_{\rm DW}$ (top), $\omega_{\rm DW}/(2 m\pi)$ (middle), $\gamma_{\rm DW}$ (bottom), shown as functions of the starting time $t_0$ of the direct-wave search. 
In all panels, the horizontal dashed lines indicate the predicted values of the corresponding quantities, computed from the maximum-posterior remnant mass and dimensionless spin inferred from the full inspiral–merger–ringdown analysis. 
In the top panel, the shaded band denotes the 90\% credible interval of $\mathscr{A}_{\rm RD}$ obtained from ringdown spectroscopy.
In the middle and the bottom panels, the error bars in red denote the 90\% credible interval of frequency and damping rate obtained in \cite{Lu:2025vol}. 
}
\label{fig:A_t}
\end{figure}

We next investigate the dependence of our inference of $\mathscr{A}_{\rm DW}$ on the choice of $t_0$.
The top panel of Fig.~\ref{fig:A_t} shows the median (markers) and 90\% credible intervals (error bars) of $p(\mathscr{A}_{\rm DW}|d, H, I)$ as a function of $t_0 \in [-7, -3],M_{\rm f}^{\rm det}$.
The horizontal dashed line indicates the horizon area using Eq.~\eqref{eq:A_RD}, assuming the median mass and spin inferred from the whole signal analysis, while the shaded band denotes the 90\% credible interval of $\mathscr{A}_{\rm RD}$, which is independent of the start time of the analysis \ref{item:S2}.
We find that for $t_0 \in [-4.5, -3] M_{\rm f}^{\rm det}$, except for $t_0 = -4M_{\rm f}^{\rm det} $, the credible interval of $p(\mathscr{A}_{\rm DW}|d, H, I)$ encloses the predicted value and significantly overlaps with that of $\mathscr{A}_{\rm RD}$.
This behavior indicates that, within this window of $t_0$, the direct-wave measurement provides an accurate probe of the horizon area.
The localized deviation near $t_0\simeq -4M_{\rm f}^{\rm det}$ does not persist at neighboring start times, suggesting that it is a start-time-dependent fluctuation rather than a systematic breakdown of the direct-wave interpretation. 
Such fluctuations can arise at a finite signal-to-noise ratio, especially because the direct-wave and quasinormal-mode components are fitted simultaneously in a short near-merger window.

For earlier start times, $t_0 \leq -4.5 M_{\rm f}^{\rm det}$, the posterior $p(\mathscr{A}_{\rm DW} | d, H, I)$ no longer encloses the predicted horizon area, nor does it significantly overlap with the 90\% credible interval of $\mathscr{A}_{\rm RD}$.
This breakdown can be traced to the behavior of the inferred direct-wave frequency and damping rate.
The second and third panels of Fig.~\ref{fig:A_t} show the median (markers) and 90\% credible intervals (error bars) of the marginalized posteriors of the frequency $\omega_{\rm DW}/(2m\pi)$ and damping rate$\gamma_{\rm DW}$, respectively.
The horizontal dashed lines indicate the predicted values computed from $M_{\rm f}^{\rm det} = 68.1 M_{\odot}$ and $a = 0.68$.
For comparison, we also show the corresponding results from \cite{Lu:2025vol} (red), which exhibit good agreement with our analysis.
At $t_0=-3M_{\rm f}^{\rm det}$, the fitted direct-wave quantities
$(\omega_{\rm DW}/m) $and $(\gamma_{\rm DW})$ are consistent with the Kerr-remnant predictions for
$\Omega_{\rm H}$ and $\kappa$, respectively, apart from the localized start-time-dependent fluctuation near $(t_0=-4M_{\rm f}^{\rm det})$.
However, as $t_0$ decreases, this agreement breaks down: for $-6.5 M_{\rm f}^{\rm det} \lesssim t_0 \lesssim -3.5 M_{\rm f}^{\rm det}$, the inferred $\Omega_{\rm H}$ becomes inconsistent with the prediction, while for $t_0 \leq -4.5 M_{\rm f}^{\rm det}$, the inferred $\kappa$ also deviates significantly.
This pattern indicates that, for $t_0 \lesssim -4.5 M_{\rm f}^{\rm det}$, the mapping between the direct-wave frequency and damping rate and the horizon quantities $\Omega_{\rm H}$ and $\kappa$ ceases to be reliable. 
For $t_0 \lesssim -4.5 M_{\rm f}^{\rm det}$, both $\Omega_{\rm H}$ and $\kappa$ are expected to undergo dynamical evolution \cite{Lu:2025vol}.
This breakdown explains the loss of accuracy in the inferred $\mathscr{A}_{\rm DW}$ at early start times.
A quantitative diagnostic of this start-time dependence is given in the End Matter.

\vspace{0.2cm}
\noindent \textit{Test of the Hawking area law.} The measurement of the black-hole horizon area enables a range of applications, among which a natural one is to test the Hawking area law.
Under the effective perturbed-Kerr interpretation described above, \(\mathscr A_{\rm DW}\) is a Kerr-equivalent area associated with the direct-wave regime, while \(\mathscr A_{\rm RD}\) is the late-time Kerr-remnant area inferred from ringdown. We therefore consider the quantity
\begin{equation}\label{eq:DeltaA}
\Delta \mathscr{A} = \mathscr{A}_{\rm RD} - \mathscr{A}_{\rm DW},
\end{equation}
which, if the area law holds, is expected to be non-decreasing, $\Delta \mathscr{A} \geq 0$.
At zeroth order in the perturbation of the remnant, the direct-wave and ringdown regimes are described by the same Kerr-equivalent area, so \(\Delta\mathscr A=0\). The first nonzero contribution is the first-law area increment, \(\delta\mathscr A\), which is expected to satisfy \(\delta\mathscr A\ge0\) under the assumptions stated above. For GW250114, however, this increment is well below the sensitivity of the present direct-wave measurement.
Using a first-law estimate of the horizon flux, we find that resolving this first-order area increment for GW250114 would require a direct-wave signal-to-noise ratio of order $10^3$ (see End Matter). 
In practice, therefore, we must restrict our analysis to the leading-order contribution of the change in the area, which is simply $\Delta \mathscr{A} = 0$ to ${\cal{O}}(\delta \mathscr{A})$. 
This is the consistency test we now focus on.

To perform the test, we construct the posterior distribution for $\Delta \mathscr{A}$, $p(\Delta \mathscr{A}|d, H, I)$, as follows.
Each posterior sample of $\Omega_{\rm H}$ and $\kappa$ from method~\ref{item:S1}, and of $M$ and $a$ from method~\ref{item:S2}, is mapped to a corresponding sample of $\mathscr{A}_{\rm DW}$ and $\mathscr{A}_{\rm RD}$, respectively.
Since the numbers of posterior samples in methods~\ref{item:S1} and \ref{item:S2} may differ, we construct $\Delta \mathscr{A}$ by randomly drawing $2\times 10^5$ samples from the sets of $\mathscr{A}_{\rm DW}$ and $\mathscr{A}_{\rm RD}$, allowing for repeated draws.
For each pair of samples, we compute $\Delta \mathscr{A} = \mathscr{A}_{\rm RD} - \mathscr{A}_{\rm DW}$.
The resulting samples are then used to approximate $p(\Delta \mathscr{A}|d, H, I)$ via kernel density estimation.

\begin{figure}[tp!]
\centering  
\subfloat{\includegraphics[width=\columnwidth]{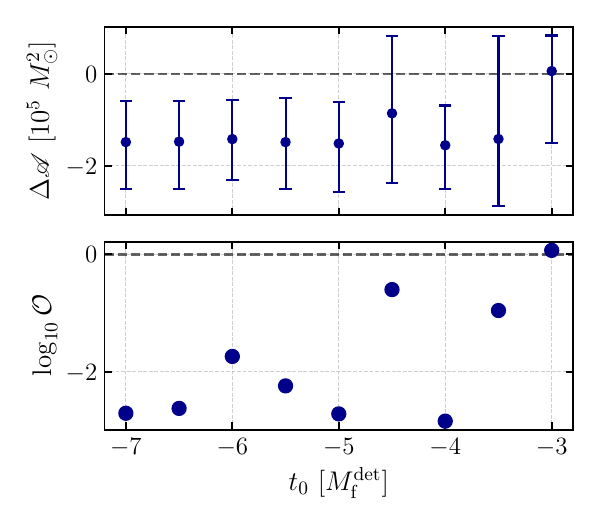}}
\caption{The median (markers) and 90\% credible intervals (error bars) of the marginalized posterior for $\Delta \mathscr{A}$ (top), and the base-10 logarithm of the odds ratio ($\log_{10} \mathcal{O}$, bottom), as functions of the starting time $t_0$ of the direct-wave search. 
}
\label{fig:Summary_plot}
\end{figure}

The black dash–dotted curve in Fig.~\ref{fig:PDF_A_m3M} shows $p(\Delta \mathscr{A}|d, H, I)$ obtained from a direct-wave search starting at $t_0 = -3 M_{\rm f}^{\rm det}$, the latest time at which direct waves have been identified for GW250114 \cite{Lu:2025vol}.
The distribution exhibits substantial support near $\Delta \mathscr{A}=0$, indicating consistency with the area law.
We also examine the dependence of $p(\Delta \mathscr{A}|d, H, I)$ on $t_0$.
The top panel of Fig.~\ref{fig:Summary_plot} shows the median (markers) and 90\% credible intervals (error bars) of $\Delta \mathscr{A}$ as functions of $t_0$, with the horizontal dashed line indicating $\Delta \mathscr{A}=0$.
We find that, for $-4.5 M_{\rm f}^{\rm det} \leq t_0 \leq -3 M_{\rm f}^{\rm det}$ (again, except at $t_0 = -4 M_{\rm f}^{\rm det} $), the credible interval encloses $\Delta \mathscr{A}=0$, consistent with the area law.
The localized deviation at $t_0=-4M_{\rm f}^{\rm det}$ follows from the same start-time-dependent fluctuation discussed below Fig.~\ref{fig:A_t}. 
For earlier start times, however, the interval no longer includes zero and the distribution shifts toward negative values.
This trend can be attributed to the behavior of the inferred direct-wave parameters: as $t_0$ decreases, the extracted $\Omega_{\rm H}$ and $\kappa$ increasingly deviate from the remnant’s horizon angular velocity and surface gravity, leading to a biased estimate of $\mathscr{A}_{\rm DW}$.
Nevertheless, within the window $t_0 \in [-4.5, -3]M_{\rm f}^{\rm det}$, the direct-wave signal still provides a reliable estimate of $\Delta \mathscr{A}$ and enables a meaningful test of the area law.

Using $p(\Delta \mathscr{A}|d, H, I)$, we compute the odds ratio
\begin{align}
\mathcal{O} =
\frac{
\displaystyle \int_{0}^{+\infty} d\Delta \mathscr{A}\, p(\Delta \mathscr{A}\mid d, H, I)
}{
\displaystyle \int_{-\infty}^{0} d\Delta \mathscr{A}\, p(\Delta \mathscr{A}\mid d, H, I)
}.
\end{align}
Values of $\mathcal{O}>1$ ($<1$) indicate that the data favor (disfavor) consistency with the Hawking area law.
As mentioned before, at leading order in perturbation theory, $\mathscr{A}_{\rm DW}$ and $\mathscr{A}_{\rm RD}$ are expected to coincide, so that $\Delta \mathscr{A}$ is a first-order quantity (that we cannot resolve) and its posterior is expected to peak near $\Delta \mathscr{A}=0$.
Accordingly, we interpret $\mathcal{O}\approx 1$ as the absence of evidence for a violation of the area law.
In contrast, a significant deviation toward $\mathcal{O}<1$, within reasonable analysis settings, would indicate evidence for a violation. A significant deviation toward $\mathcal{O}>1$ would imply a measurement of the first-order correction $\delta  \mathscr{A}$, which would require a much higher signal-to-noise ratio. 

The second panel of Fig.~\ref{fig:Summary_plot} shows $\log_{10}\mathcal{O}$ as a function of $t_0$.
We find that $\log_{10}\mathcal{O}\approx 0$ for $-4.5 M_{\rm f}^{\rm det} \leq t_0 \leq -3 M_{\rm f}^{\rm det} $ (except for $t_0 = -4 M_{\rm f}^{\rm det}$), corresponding to $\mathcal{O}\sim 1$ and indicating no strong evidence for a violation of the Hawking area law. 
For earlier start times, $\log_{10}\mathcal{O}$ decreases significantly, reaching values of $\sim -3$ for $-7 M_{\rm f}^{\rm det} \leq t_0 \leq -5 M_{\rm f}^{\rm det}$.
This behavior mirrors the shift of the posterior of $\Delta \mathscr{A}$ toward negative values (c.f. the top panel of Fig.~\ref{fig:Summary_plot}).
However, this same range of $t_0$ corresponds to the regime in which $\mathscr{A}_{\rm DW}$ is no longer consistent with the predicted horizon area, indicating a breakdown of the near-horizon interpretation of the direct-wave parameters.
We therefore attribute the apparent preference for $\mathcal{O}<1$ in this region to modeling bias rather than to genuine evidence for a violation of the area law.
Taken together, these results show that, within $-4.5 M_{\rm f}^{\rm det} \lesssim t_0 \lesssim -3 M_{\rm f}^{\rm det}$, the data remain consistent with the Hawking area law, in agreement with the behavior of $\Delta \mathscr{A}$ in the top panel.

\vspace{0.2cm}

\noindent \textit{Conclusions.} We have introduced a method to infer a Kerr-equivalent black-hole horizon area from direct waves in binary black-hole coalescences. The method interprets the fitted direct-wave frequency and damping rate as effective estimates of the horizon angular velocity and surface gravity of a perturbed post-merger Kerr black hole, and maps them to a Kerr-equivalent area.
Applied to GW250114, we find that, within the direct-wave windows, the inferred area is consistent with that predicted from the late-time remnant mass and spin. 
The consistency between the direct-wave- and ringdown-based estimates is itself an important result: it indicates that, once the near-horizon regime is reached, direct waves provide an independent probe of remnant horizon quantities beyond the standard quasinormal-mode framework.
As a first application, we perform a conditional near-merger test of Hawking's area law and find no strong evidence for violations in the regime where the direct-wave area estimator is self-consistent.

Beyond this application, the direct-wave measurement of the horizon area offers new opportunities for astrophysics.
In particular, it provides an additional constraint on the remnant spin, which is typically less well-measured than the mass. 
This may improve population inference of black-hole mergers \cite{LIGOScientific:2025pvj} and our understanding of the formation channels of black holes of mass of $(10$--$200) M_{\odot}$ \cite{Antonelli:2023gpu, Fishbach:2017dwv}.

Our analysis also independently supports the recent identification of direct waves in GW250114 \cite{Lu:2025vol}.
Our approach models and subtracts quasinormal modes within a waveform framework, whereas~\cite{Lu:2025vol} employs rational filters. 
Despite methodological differences, the inferred frequency and damping rate are in good agreement, providing a robust cross-check of the direct-wave signal.

Looking ahead, several directions merit further investigation.
In particular, relaxing the assumption of fixed remnant parameters would improve the consistency of the measurement, and a more precise understanding of the relation between direct-wave properties and horizon quantities could further enhance the accuracy of the inferred area.
Overall, direct-wave measurements of black-hole horizon area open a new avenue for probing strong-gravity physics and black-hole astrophysics with gravitational waves.

\appendix

\vspace{0.2cm}
\noindent \textit{Acknowledgments.}  We thank the authors of \cite{Lu:2025vol} for sharing their direct-wave search results with us for comparison.  
The authors would like to thank Sizheng Ma for detailed comments on the initial version of the manuscript. 
We also thank Yanbei Chen, Nicholas Manton, and Ulrich Sperhake for insightful discussion. 
A. K. W. C.~acknowledges the Herchel Smith Fellowship at the University of Cambridge for support of this work.
K. K. H. L., A. L. and N. Y. acknowledge support from the
Simons Foundation (via Award No. 896696), the Simons
Foundation International (via Grant No. SFI-MPS-
BH-00012593-01), and the NSF (via Grants No. PHY-
2512423).
The calculations and results reported in this work were produced using the computational resources of the Illinois Campus Cluster, which is a computing resource that is operated by the Illinois Campus Cluster Program (ICCP) in conjunction with National Center for Supercomputing Applications (NCSA), and is supported by funds from the University of Illinois at Urbana-Champaign. 
This work made use of \texttt{astropy}~\citep{astropy2013,astropy2018,astropy2022}, 
\texttt{h5py}~\citep{collette2013python}, \texttt{matplotlib}~\citep{hunter2007matplotlib}, \texttt{numpy}~\citep{harris2020numpy}, and \texttt{scipy}~\citep{virtanen2020scipy}.
This material is based upon work supported by NSF's LIGO Laboratory which is a major facility fully funded by the National Science Foundation.
This manuscript carries a LIGO document number of P2600278. 


\vspace{0.2cm}
\noindent \textit{Data availability.}  The gravitational-wave strain data analyzed in this work are publicly available through the Gravitational Wave Open Science Center (GWOSC) at \url{https://gwosc.org/eventapi/html/O4_Discovery_Papers/GW250114_082203/v1/}. 
The time-domain inference package \texttt{pyRing}, used in this work to search for direct waves, is publicly available at \url{https://git.ligo.org/lscsoft/pyring}.

\section*{End Matter}

\begin{figure}[tp!]
\centering  
\subfloat{\includegraphics[width=\columnwidth]{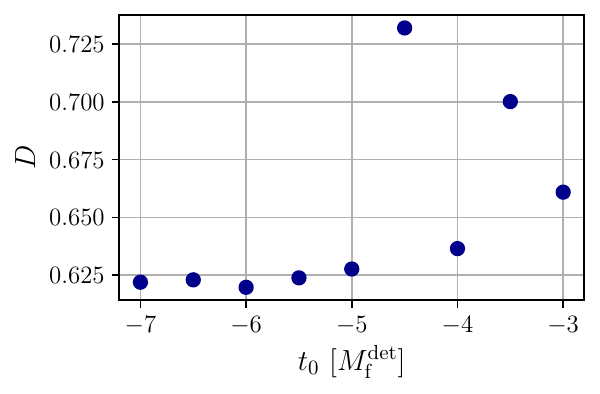}}
\caption{The diagnostic measures the posterior-level fractional deviation of the fitted direct-wave quantities
$[\omega_{\rm DW}/(2m \pi),\gamma_{\rm DW}]$ from the Kerr-remnant predictions
$[\Omega_{\rm H}/(2\pi),\kappa]$, and includes both posterior width and posterior displacement.
}
\label{fig:Divergence}
\end{figure}

\vspace{0.2cm}
\noindent \textit{Start-time diagnostic.} To quantify the breakdown of the correspondence \(\gamma_{\rm DW}\simeq\kappa\) and \(\omega_{\rm DW}\simeq m\Omega_{\rm H}\), we define the divergence
\begin{equation}\label{eq:Divergence}
D = \sqrt{\sum_{i=1,2}\left\langle\left(1 - \frac{\theta_{i}}{\bar{\theta}_{i}} \right)^2\right\rangle},
\end{equation}
where $\theta_1 = \omega_{\rm DW}/(2 m \pi)$, $\theta_2 = \gamma_{\rm DW}$, and $\bar{\theta}_{1,2}$ are their values predicted assuming $\omega_{\rm DW} = m \Omega_{\rm H}$ and $\gamma_{\rm DW} = \kappa$, and assuming the maximal-posterior values of $M$ and $a$ obtained from the whole-signal analysis.
In Eq.~\eqref{eq:Divergence}, \(D\) is normalized so that it is dimensionless.
Here,
\begin{equation}
\left\langle\left(1-\frac{\theta_{i}}{\bar{\theta}_{i}} \right)^2\right\rangle = \int d\theta_{i} \left(1-\frac{\theta_{i}}{\bar{\theta}_{i}} \right)^2 p(\theta_{i}|d, H, I),
\end{equation}
with $p(\theta_{i}|d, H, I)$ the marginalized posterior of $\theta_{i}$. 
Figure~\ref{fig:Divergence} shows $D$ as a function of the direct-wave search start time.
Across the start times considered here, $D$ remains of order unity and varies moderately with $t_0$.
Assuming comparable contributions from the $\Omega_{\rm H}$ and $\kappa$ terms, $D/\sqrt{2}$ gives the typical RMS fractional deviation of each parameter from its Kerr-remnant prediction.
The values shown in Fig.~\ref{fig:Divergence} therefore correspond to typical fractional deviations of $\sim 40\%$--$50\%$, consistent with the widths and displacements of the $\Omega_{\rm H}$ and $\kappa$ posteriors shown in Fig.~\ref{fig:Summary_plot}.
The localized variations of $D$ support the interpretation that the corresponding area deviations arise from start-time-dependent fluctuations in the direct-wave--ringdown decomposition, rather than from a systematic violation of the area law.

\vspace{0.2cm}
\noindent \emph{Sensitivity to the first-order area increment.---}
Here we estimate the direct-wave signal-to-noise ratio required to resolve the first-order area increment expected from the black-hole first law. 
For perturbations between nearby stationary black-hole states, we write
\begin{equation}
\delta \mathscr{A}
=
\frac{8\pi}{\kappa}
\left(\delta E-\Omega_{\rm H}\delta J\right)
\equiv
\frac{8\pi}{\kappa} Q_{\rm H},
\end{equation}
where \(Q_{\rm H}\equiv \delta E-\Omega_{\rm H}\delta J\). 
For radiation dominated by an azimuthal mode \(m\) with angular frequency \(\omega\), one has approximately \(\delta J/\delta E\simeq m/\omega\), and therefore
\begin{equation}
Q_{\rm H}
\simeq
\delta E
\left(
1-\frac{m\Omega_{\rm H}}{\omega}
\right).
\end{equation}
For the GW250114 remnant parameters used in this work, \(M_{\rm f}^{\rm det}=68.1M_\odot\) and \(a=0.68\), the quadrupolar superradiant frequency is \(2\Omega_{\rm H}/(2\pi)\simeq186\,{\rm Hz}\). 
Thus, if the direct-wave frequency near peak amplitude lies close to \(m\Omega_{\rm H}\), the first-law combination \(Q_{\rm H}\), and hence the corresponding area increment, is suppressed.

We estimate the measurement precision required to resolve this increment by propagating the direct-wave uncertainties in \(\Omega_{\rm H}\) and \(\kappa\) through Eq.~\eqref{eq:A_DW}. 
Writing \(R=\sqrt{\Omega_{\rm H}^{2}+\kappa^{2}}\), the fractional area uncertainty is, to linear order,
\begin{equation}
\frac{\sigma_{\mathscr{A}}}{\mathscr{A}}
\simeq
\sqrt{
C_{\Omega}^{2}\epsilon_{\Omega}^{2}
+
C_{\kappa}^{2}\epsilon_{\kappa}^{2}
},
\end{equation}
where \(\epsilon_{\Omega}\equiv\sigma_{\Omega_{\rm H}}/\Omega_{\rm H}\), 
\(\epsilon_{\kappa}\equiv\sigma_{\kappa}/\kappa\), and
\begin{align}
C_{\Omega}
&\equiv
\frac{\partial \ln \mathscr{A}}{\partial \ln \Omega_{\rm H}}
=
-\frac{\Omega_{\rm H}^{2}}{R^{2}}
-\frac{\Omega_{\rm H}^{2}}{R(R+\kappa)},\\
C_{\kappa}
&\equiv
\frac{\partial \ln \mathscr{A}}{\partial \ln \kappa}
=
-\frac{\kappa^{2}}{R^{2}}
-\frac{\kappa}{R+\kappa}
\left(\frac{\kappa}{R}+1\right).
\end{align}
For \(a=0.68\), these coefficients are \(C_{\Omega}\simeq -0.73\) and \(C_{\kappa}\simeq -1.27\), showing that the inferred area is more sensitive to fractional errors in \(\kappa\) than to comparable fractional errors in \(\Omega_{\rm H}\), which is also consistent with our observation that the breakdown of the correspondence of $\gamma_{\rm DW} \simeq \kappa$ in an earlier $t_0$ leads to inaccuracy in measuring $\mathscr{A}_{\rm DW}$. 

Assuming the usual inverse-SNR scaling of parameter uncertainties,
\begin{equation}
\frac{\sigma_{\mathscr{A}}}{\mathscr{A}}(\rho_{\rm DW})
\simeq
\left(\frac{\rho_{0}}{\rho_{\rm DW}}\right)
\left(\frac{\sigma_{\mathscr{A}}}{\mathscr{A}}\right)_{0},
\end{equation}
where \(\rho_{\rm DW}\) is the direct-wave SNR and \(\rho_{0}\) denotes the present direct-wave SNR, resolving a target first-order fractional area increment \(\Delta_{\mathscr{A}}\equiv |\delta\mathscr{A}|/\mathscr{A}_{\rm f}\) at \(n\sigma\) requires
\begin{equation}
\rho_{\rm DW}
\gtrsim
n\rho_{0}
\frac{(\sigma_{\mathscr{A}}/\mathscr{A})_{0}}
{\Delta_{\mathscr{A}}}.
\end{equation}
Using the present direct-wave measurement as normalization gives a required direct-wave SNR of order \(10^{3}\) for GW250114-like parameters. 
The precise value depends sensitively on how close the direct-wave frequency near peak amplitude is to the superradiant frequency \(m\Omega_{\rm H}\), but the conclusion is robust: current data are sensitive to the leading-order consistency condition \(\Delta\mathscr{A}\simeq0\), rather than to the small positive first-order area increase itself.

\bibliography{ref}

\end{document}